# On the Meissner effect in SU(2) lattice gauge theory

P. Cea [a] and L. Cosmai[*] [b]

[a]Dipartimento di Fisica dell'Università di Bari, 70126 Bari, Italy

[b] Istituto Nazionale di Fisica Nucleare, Sezione di Bari, 70126 Bari, Italy

We investigate the dual superconductor model of color confinement in SU(2) lattice gauge theory. We find that the transverse distribution of the longitudinal chromoelectric field between static quarks displays the dual Meissner effect. We also give evidence that the problem of color confinement could be approached in the framework of the 't Hooft's Abelian projection.

To understand the non perturbative phenomenon of the color confinement G. 't Hooft[1] and S. Mandelstam[2] proposed a model known as *dual superconductor model*. The physical concepts of this model stem from the theory of (electric) superconductivity.

In 't Hooft's formulation the dual superconductor model is elaborated in the framework of the Abelian projection[3]. After a particular gauge has been fixed, reducing the symmetry to that of the maximal Abelian (Cartan) subgroup, the non Abelian gauge theory is described in terms of Abelian projected gauge fields ("photons"). In this scenario there are also color magnetic monopoles whose condensation should cause the confinement of all particles which are color electrically charged with respect to the above "photons".

We direct our analysis to SU(2) lattice gauge theory. In particular we want to investigate the dual Meissner effect. To this purpose we study on the lattice the transverse distribution of the longitudinal chromoelectric field $E_l(x_\perp)$ between a static $q\bar{q}$ pair. We first measure $E_l$ on SU(2) lattice gauge configurations. $E_l$ is proportional to the connected correlation function[4] of the plaquette $U_P$ with the parallel Wilson loop $W$ connected by the Schwinger line $L$

$$\rho_W = \frac{\langle \text{tr}\left(WLU_PL^\dagger\right)\rangle}{\langle \text{tr}(W)\rangle} - \frac{1}{2}\frac{\langle \text{tr}(U_P)\text{tr}(W)\rangle}{\langle \text{tr}(W)\rangle} \; . \quad (1)$$

By varying the distance of the plaquette $U_P$ to the Wilson loop $W$ we are able to scan the field $E_l(x_\perp)$.

---
[*]Talk presented by L. Cosmai

If the chromomagnetic vacuum behaves like a type II superconductor, the Meissner effect gives rise to Abrikosov vortices. If the penetration length $\lambda$ is much bigger than the coherence length $\xi$, the solution of the dual London equation is

$$E_l(x_\perp) = \frac{\Phi_e}{2\pi}\frac{1}{\lambda^2}K_0\left(\frac{x_\perp}{\lambda}\right) \; , \quad x_\perp > 0 \; . \quad (2)$$

where $\Phi_e$ is the electric flux and $K_0$ is the modified Bessel function of order zero. Equation (2) is a straightforward consequence of the dual superconductor hypothesis (see also Refs. [5,6]).

We extract $E_l$ from correlations with $6 \times 6$ or $4 \times 8$ Wilson loops on a $16^4$ lattice. Figure 1 shows the behavior of $E_l$ vs. transverse distance $x_\perp$ with superimposed the fit (solid line)

$$E_l(x_\perp) = A_M K_0\left(\mu_M x_\perp\right) \; , \quad x_\perp > 0 \; . \quad (3)$$

which derives from Eq.(2). The quality of the fit improves if we discard the point at distance 1 (dashed line); this is reasonable since Eq.(3) applies in the region $\lambda \gg \xi$. The parameters $A_M$ and $\mu_M$ extracted from the fit are consistent if we use $6 \times 6$ Wilson loop or $4 \times 8$ Wilson loop in Eq.(1).

We have also studied the chromoelectric field $E_l$ obtained by measuring the connected correlation function Eq.(1) built out from Abelian projected links in the maximal Abelian gauge:

$$\rho_W^{ab} = \frac{\langle \text{tr}\left(W^A U_P^A\right)\rangle}{\langle \text{tr}\left(W^A\right)\rangle} - \frac{1}{2}\frac{\langle \text{tr}\left(U_P^A\right)\text{tr}\left(W^A\right)\rangle}{\langle \text{tr}\left(W^A\right)\rangle} \; . \quad (4)$$

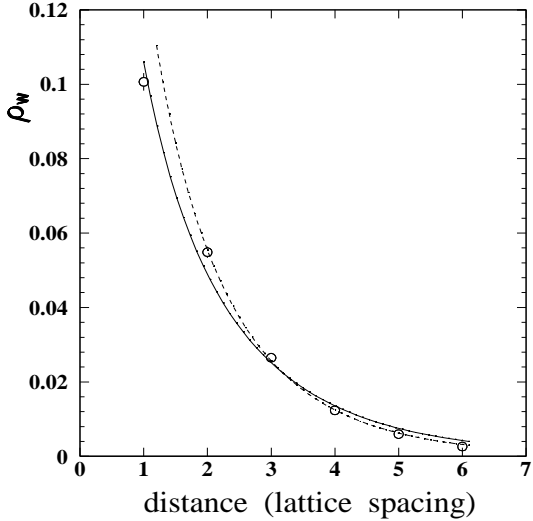

Figure 1. Transverse distribution of the longitudinal chromoelectric field at $\beta = 2.5$. Solid and dashed lines are discussed in the text.

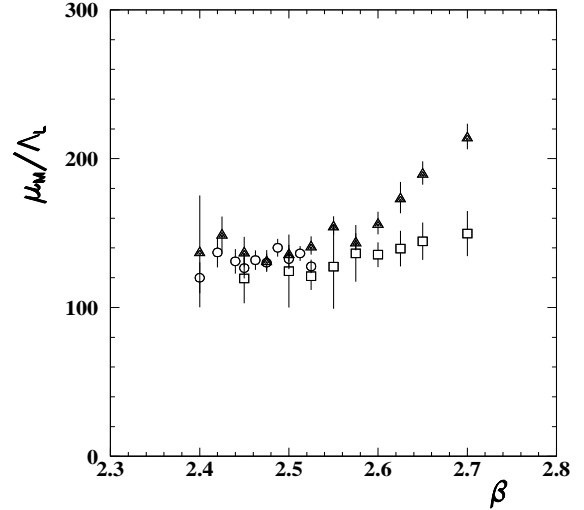

Figure 2. $\mu_m/\Lambda_L$ vs. $\beta$. Values from Abelian projected $12^4$ lattice ($\bigcirc$), Abelian projected $16^4$ lattice ($\square$), and SU(2) $16^4$ lattice ($\blacktriangle$).

The superscript means that the Wilson loop and the plaquette are built out from the Abelian projected links. Maximal Abelian gauge on the lattice[7] is fixed by means of an iterative overrelaxed algorithm.

The penetration length $\lambda = 1/\mu_M$ is obtained by fitting Eq.(3) to our numerical data. Figure 2 displays the values of the parameter $\mu_M$ in units of $\Lambda_L$ vs. $\beta$ (some of the data refers to our previous results on a $12^4$ lattice[8]) obtained from Abelian projected configurations in the maximal Abelian gauge, together with $\mu_M$ obtained from SU(2) configurations. One can see that within statistical errors both measurements are in agreement up to $\beta \simeq 2.6$ where presumedly lattice artefacts become to defile numerical results.

As concern the normalization $A_M$ in Eq.(3) we observe that Eq.(2) and Eq.(3) implies that $A_M \sim \mu_M^2$. As a matter of fact Figure 3 shows that $A_M/\mu_M^2$ is approximately constant. To understand the different numerical values of the normalization extracted from SU(2) configurations ($A_M$) or from Abelian projected configurations in the maximal abelian gauge ($A_M^{ab}$) we point out that Eq.(2) implies electric flux $\Phi_e \sim$ electric charge and therefore $A_M/A_M^{ab} \sim 3$ which is the ratio between the SU(2) charge and the Abelian projected one. Actually we find out that this ratio is quite consistent with 3. However better statistics and further study is needed to fix this point.

As a further check of the dual superconductor model we measure the mass extracted from the connected correlation functions of an operator with the quantum numbers of the photon[9], built up using Abelian projected links in the maximal Abelian gauge. The photon mass $m_\gamma$ should be related to the penetration length of the chromoelectric field, i.e. $m_\gamma = 1/\lambda = \mu_M$. Figure 4 displays the ratio $\mu_M/m_\gamma$ which is consistent

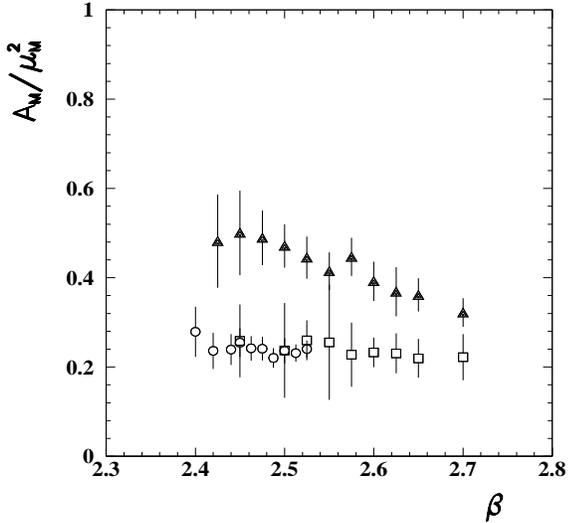

Figure 3. The ratio $A_M/\mu_M^2$ versus $\beta$. Symbols as in Fig.2.

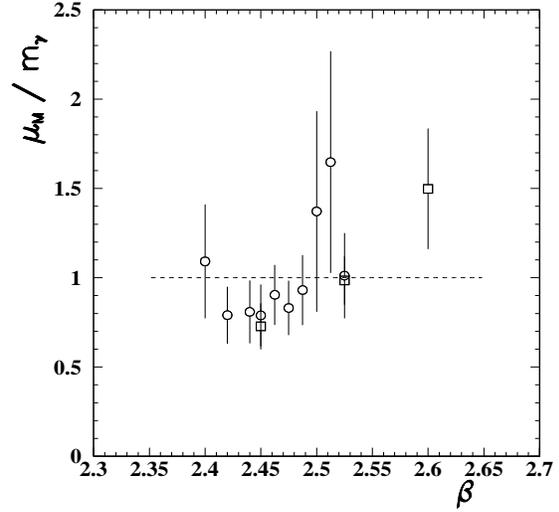

Figure 4. The ratio $\mu_M/m_\gamma$ versus $\beta$. Symbols as in Fig.2.

with 1. Large statistical fluctuations are essentially due to the extimation of the photon mass.

In conclusion we have showed that the transverse distribution of the longitudinal chromoelectric field due to static quark-antiquark pair satisfies the dual London equation. We have also given evidence that the penetration length obtained from Abelian projected links in maximal Abelian gauge is compatible with the penetration length derived from SU(2) gauge configurations without gauge fixing. Moreover we verified that the ratio of the penetration length to the photon mass is nearly 1. Thus we feel that the problem of color confinement could be approached in the framework of the dual superconductivity using the 't Hooft's Abelian projection. However further studies are needed to understand what is the analogous of the Cooper condensation which gives rise to the chromomagnetic superconducting vacuum.